\documentclass[3p,twocolumn]{elsarticle}
\pdfoutput=1 

\usepackage{graphicx}
\usepackage{lineno}
\usepackage[hyperindex,breaklinks]{hyperref}

\usepackage{atlasphysics}

\usepackage{preprintcover}  % load package
\PreprintCoverPaperTitle{Measurement with the ATLAS  detector of multi-particle azimuthal correlations in $p+$Pb collisions at $\sqrt{s_{\mathrm{NN}}}=5.02$~\TeV }
\PreprintIdNumber{CERN-PH-EP-2013-029}  % CERN preprint number
\PreprintCoverAbstract{In order to study further the long-range correlations (``ridge'') observed recently in $p+$Pb collisions at $\sqrt{s_{\mathrm{NN}}}=5.02$~\TeV, the second-order azimuthal anisotropy parameter of charged particles, $v_2$, has been measured with the cumulant method using the ATLAS detector at the LHC. In a data sample corresponding to an integrated luminosity of approximately $1~\mu \mathrm{b}^{-1}$, the parameter $v_2$ has been obtained using two- and four-particle cumulants over the pseudorapidity range $|\eta|<2.5$. The results are presented as a function of transverse momentum and the event activity, defined in terms of the transverse energy summed over $3.1<\eta<4.9$ in the direction of the Pb beam. They show features characteristic of collective anisotropic flow, similar to that observed in Pb+Pb collisions. A comparison is made to results obtained using two-particle correlation methods, and to predictions from hydrodynamic models of $p+$Pb collisions. Despite the small transverse spatial extent of the $p+$Pb collision system, the large magnitude of $v_2$ and its similarity to hydrodynamic predictions provide additional evidence for the importance of final-state effects in $p+$Pb reactions.
}  % paper abstract
\PreprintJournalName{Physics Letters B}  % journal name

\biboptions{sort&compress}

\begin{document}

\linenumbers

\title{Measurement with the ATLAS detector of multi-particle azimuthal correlations in $p+$Pb collisions at $\sqrt{s_{\mathrm{NN}}}=5.02$~\TeV}

% Author

\author{The ATLAS Collaboration}

% Abstract
\begin{abstract}
In order to study further the long-range correlations (``ridge'') observed recently in $p+$Pb collisions at $\sqrt{s_{\mathrm{NN}}}=5.02$~\TeV, the second-order azimuthal anisotropy parameter of charged particles, $v_2$, has been measured with the cumulant method using the ATLAS detector at the LHC. In a data sample corresponding to an integrated luminosity of approximately $1~\mu \mathrm{b}^{-1}$, the parameter $v_2$ has been obtained using two- and four-particle cumulants over the pseudorapidity range $|\eta|<2.5$. The results are presented as a function of transverse momentum and the event activity, defined in terms of the transverse energy summed over $3.1<\eta<4.9$ in the direction of the Pb beam. They show features characteristic of collective anisotropic flow, similar to that observed in Pb+Pb collisions. A comparison is made to results obtained using two-particle correlation methods, and to predictions from hydrodynamic models of $p+$Pb collisions. Despite the small transverse spatial extent of the $p+$Pb collision system, the large magnitude of $v_2$ and its similarity to hydrodynamic predictions provide additional evidence for the importance of final-state effects in $p+$Pb reactions.
\end{abstract}

\maketitle

\section{Introduction}
Recent observations of ridge-like structures in the two-particle correlation functions measured in proton-lead ($p+$Pb) collisions at 5.02~TeV \citep{ridge_cms,alice_ridge,ridge_atlas} have led to differing theoretical explanations. These structures have been attributed either to mechanisms that emphasize initial-state effects, such as the saturation of parton distributions in the Pb-nucleus \citep{arbuzow,dusling1,dusling2,kovchegov}, or to final-state effects, such as jet--medium interactions \citep{wong}, interactions induced by multiple partons \citep{strikman,alderweir,Ryskin,Diehl}, and collective anisotropic flow \citep{avsar,werner,deng,avsar1,bozek1,bozek2}.

The collective flow of particles produced in nuclear collisions, which manifests itself as a significant anisotropy in the plane perpendicular to the beam direction, has been extensively studied in heavy-ion experiments at the LHC \citep{Alice1,Alice2,Atlas1,atlas_2p,Cms1,Cms2} and RHIC (for a review see Refs.~\citep{flowReview,flowReview1}).
In $p+$Pb collisions the small size of the produced system compared to the mean free path of the interacting constituents might have been expected to generate weaker collective flow, if any, compared to heavy-ion collisions.

However, two-particle correlation studies performed recently on data from $p+$Pb collisions at $\sqrt{s_{\mathrm{NN}}}=5.02$~\TeV\, revealed the presence of a ``ridge'', a structure extended in the relative pseudorapidity, $\Delta\eta$, while narrow in the relative azimuthal angle, $\Delta\phi$, on both the near-side ($\Delta\phi \sim 0$) \cite{ridge_cms} and away-side ($\Delta\phi \sim \pi$) \citep{alice_ridge,ridge_atlas}. Furthermore, it was shown in Refs.~\cite{alice_ridge,ridge_atlas} that, after subtracting the component due to momentum conservation,  the $\Delta\phi$ distribution in high-multiplicity interactions exhibits a predominantly $\cos(2\Delta\phi)$ shape, resembling the elliptic flow modulation of the $\Delta\phi$ distributions in Pb+Pb collisions.

 The final-state anisotropy is usually characterized by the coefficients, $v_n$, of a Fourier decomposition of the event-by-event azimuthal angle distribution of produced particles \citep{flowReview,Poskanzer:1998yz}:
\begin{equation}
v_n = \langle \cos{n(\phi - \Psi_n)}\rangle,
\label{eq:vn}
\end{equation}
\noindent 
where $\phi$ is the azimuthal angle of the particle, $\Psi_n$ is the event-plane angle for the $n$-th harmonic, and the outer brackets denote an average over charged particles in an event.
In non-central heavy-ion collisions, the large and dominating $v_2$ coefficient is associated mainly with the elliptic shape of the nuclear overlap, and $\Psi_2$ defines the direction which nominally points in the direction of the classical impact parameter. In practice, initial-state fluctuations can blur the relationship between $\Psi_2$ and the impact parameter direction in nucleus-nucleus collisions. In contrast, $\Psi_2$ in proton-nucleus would be unrelated to the impact parameter and determined by the initial-state fluctuations. In nucleus-nucleus collisions, the $v_2$ coefficient in central collisions and the other $v_n$ coefficients in all collisions are related to various geometric configurations arising from fluctuations of the nucleon positions in the overlap region \cite{roland}.

In this Letter, a direct measurement of the second-order anisotropy parameter, $v_2$, is presented for $p+$Pb collisions at $\sqrt{s_{\mathrm{NN}}}=5.02$~\TeV. The cumulant method \citep{Borghini_first,Borghini_second,Cumulants,QC} is applied to derive $v_2$ using two- and four-particle cumulants. The cumulant method has been developed to characterize true multi-particle correlations related to the collective expansion of the system, while suppressing correlations from resonance decays, Bose--Einstein correlations and jet production. Emphasis is placed on the estimate of $v_2$, $v_2\{4\}$, obtained from the four-particle cumulants which are expected to be free from the effects of short-range two-particle correlations, e.g. from resonance decays, unlike the two-particle cumulants, used to estimate $v_2\{2\}$.

The measurements of multi-particle cumulants presented in this Letter should provide further constraints on the origin of long-range correlations observed in $p+$Pb collisions.

\section{Event and track selections}
The $p+$Pb data sample was collected during a short run in September 2012, when the LHC delivered $p+$Pb collisions at the nucleon--nucleon centre-of-mass energy $\sqrt{s_{\mathrm{NN}}}$ = 5.02 TeV with the centre-of-mass frame shifted by $-0.47$ in rapidity  relative to the nominal ATLAS coordinate frame\footnote{ATLAS uses a right-handed coordinate system with its origin at
the nominal interaction point (IP) in the centre of the detector
and the $z$-axis along the beam pipe. The $x$-axis points from the
IP to the centre of the LHC ring, and the $y$-axis points upward.
Cylindrical coordinates ($r,\phi$) are used in the transverse plane, $\phi$
being the azimuthal angle around the beam pipe. For the $p+$Pb
collisions, the incident Pb beam travelled in the $+z$ direction.
The pseudorapidity is defined in laboratory coordinates in terms
of the polar angle $\theta$ as $\eta = -\ln\tan(\theta/2)$. Transverse momentum and energy are defined as $\pT = p \sin\theta$ and $E_{\mathrm{T}} = E \sin\theta$, respectively.}.

The measurements were performed using the ${\mathrm{ATLAS}}$ detector \cite{atlas_det}. The inner detector (ID) was used for measuring trajectories and momenta of charged particles for $|\eta|<2.5$ with the silicon pixel detector and silicon microstrip detectors (SCT), and a transition radiation tracker, all placed in a 2~T axial magnetic field. For event triggering, two sets of Minimum Bias Trigger Scintillators (MBTS), located symmetrically in front of the endcap calorimeters, at $z=\pm3.6$~m and covering the pseudorapidity range $2.1 < |\eta| <3.9$, were used. The trigger used to select minimum-bias $p+$Pb collisions requires a signal in at least two MBTS counters. This trigger is fully efficient for events with more than four reconstructed tracks with $p_{\mathrm{T}}>0.1$~GeV. The forward calorimeters (FCal), consisting of two symmetric systems with tungsten and copper absorbers and liquid argon as the active material, cover $3.1 < |\eta| <4.9$ and are used to characterize the overall event activity.

The event selection follows the same requirements as used in the recent two-particle correlation analysis \cite{ridge_atlas}.  Events are required to have a reconstructed vertex with its $z$ position within $\pm150$~mm of the nominal interaction point. Beam--gas and photonuclear interactions are suppressed by requiring at least one hit in a MBTS counter on each side of the interaction point and at most a 10~ns difference between times measured on the two sides to eliminate through-going particles. To eliminate multiple $p+$Pb collisions (about 2\% of collision events have more than one reconstructed vertex), the events with two reconstructed vertices that are separated in $z$ by more than 15~mm are rejected.  In addition, for the cumulant analysis presented here, it is required that the number of reconstructed tracks per event, passing the track selections as described below, is greater than three. With all the above selections, the analysed sample consists of about $1.9 \times 10^6$ events.

Charged particle tracks are reconstructed in the ID using the standard algorithm optimized for $p+p$ minimum-bias measurements \cite{pptracking}.  Tracks are required to have at least six hits in the SCT detector and at least one hit in the pixel detector. A hit in the first pixel layer is also required when the track crosses an active pixel module in that layer. Additional requirements are imposed on the transverse ($d_0$) and longitudinal ($z_0\sin\theta$) impact parameters measured with respect to the primary vertex. These are: $|d_0|$ and $|z_0\sin\theta|$ must be smaller than 1.5~mm and must satisfy $|d_0/\sigma_{d_0}|<3$ and $|z_0\sin\theta/\sigma_z|<3$, where $\sigma_{d_0}$ and $\sigma_z$ are uncertainties on the transverse and longitudinal impact parameters, respectively, as obtained from the covariance matrix of the track fit. The analysis is restricted to charged particles with $0.3 < p_{\mathrm{T}} < 5.0$~GeV and $|\eta|<2.5$. 

\begin{figure}[t]
\begin{center}
\includegraphics[width=65mm]{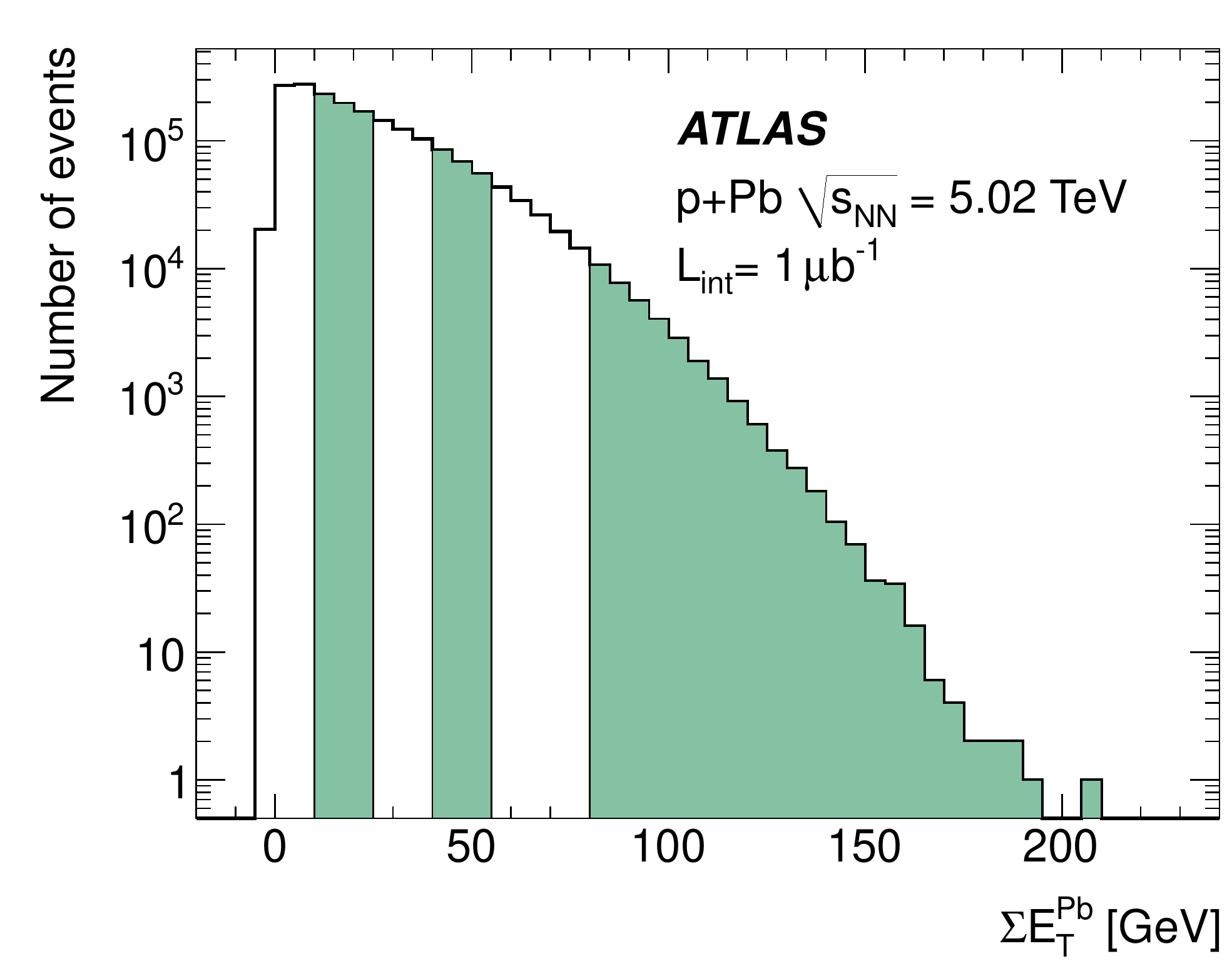}
\includegraphics[width=67mm]{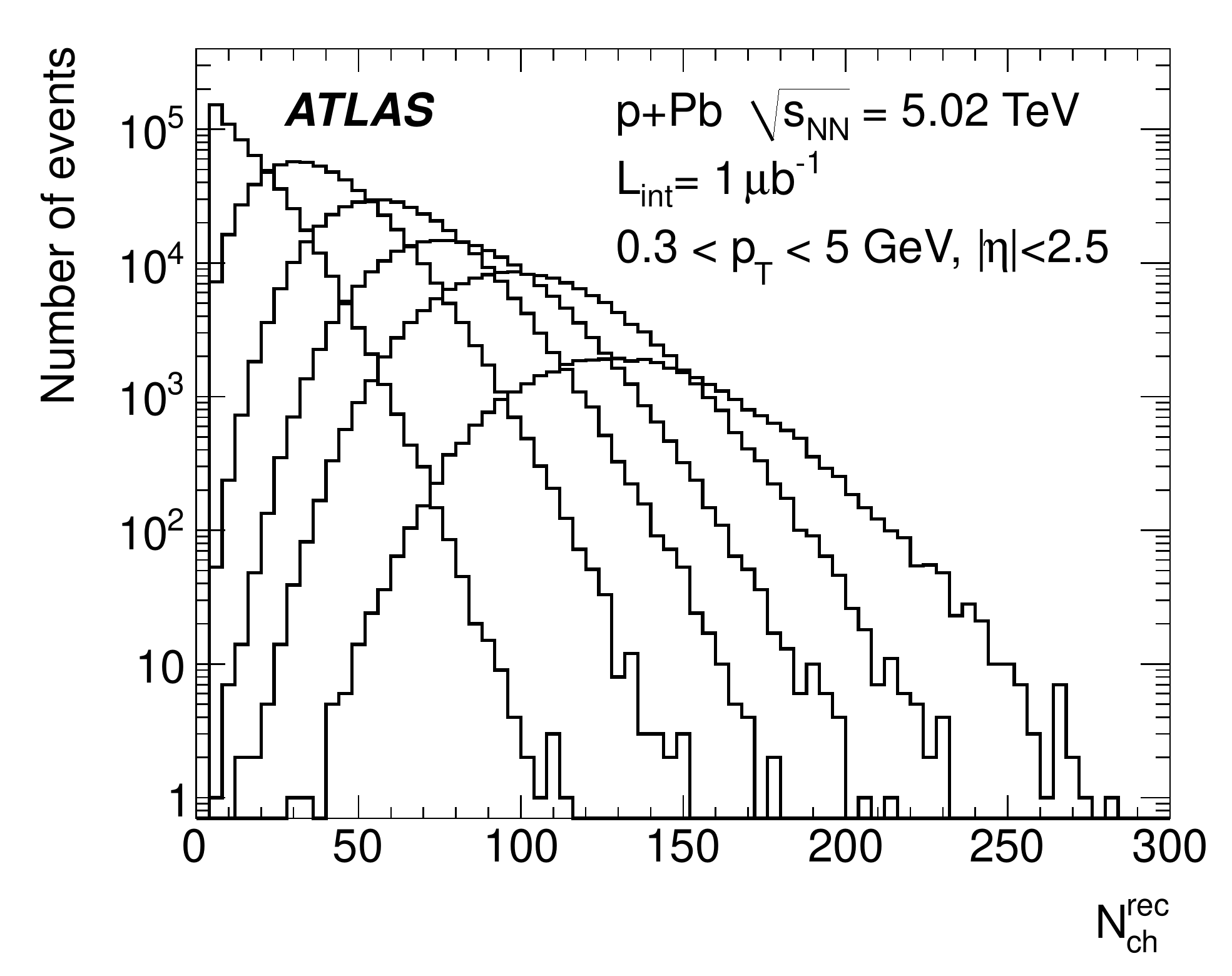}
\caption[centrality  plots]
{Upper plot: the $\Sigma E^{\mathrm{Pb}}_{\mathrm{T}}$ distribution with the six activity intervals indicated. Lower plot: the distribution of $N_{\mathrm{ch}}^{\mathrm{rec}}$ for each activity  interval. The leftmost distribution corresponds to the interval with the lowest $\Sigma E^{\mathrm{Pb}}_{\mathrm{T}}$, etc.}
\label{fig:EtNtrk}
\end{center}
\end{figure}
 \begin{table}[t]
  \begin{center}
    \begin{tabular}{|c|c|c|c|}
    \hline
     $\Sigma E^{\mathrm{Pb}}_{\mathrm{T}}$ & $\langle \Sigma E^{\mathrm{Pb}}_{\mathrm{T}} \rangle$ & range in  & $\langle N_{\mathrm{ch}}^{\mathrm{rec}} \rangle$  \\
     range & & fraction & (RMS) \\

     [GeV] & [GeV] & of events [\%] &  \\ \hline \hline
      $> 80$ & 93.7 &  0--1.9 & 134 (31) \\ \hline
      55--80 & 64.8 &  1.9--9.1 & 102 (26) \\ \hline
      40--55 & 46.7 &  9.1--20.0 & 80 (23) \\  \hline
      25--40 & 31.9 &  20.0--39.3 & 60 (20) \\ \hline
      10--25 & 16.9 & 39.3--70.4 & 37 (17) \\ \hline
      $< 10$ & 4.9 &  70.4--100 & 16 (11) \\ \hline
    \end{tabular}
    \caption[centEt]{Characterization of activity intervals as selected by $\Sigma E^{\mathrm{Pb}}_{\mathrm{T}}$. In the last column, the mean and RMS of the number of reconstructed charged particles with $|\eta|<2.5$ and $0.3 < \pT <5$~GeV, $N_{\mathrm{ch}}^{\mathrm{rec}}$, is given for each activity interval. }
    \label{tab:centET}
  \end{center}
\end{table}
The tracking efficiency is evaluated using HIJING-generated \cite{hijing} $p+$Pb events that are fully simulated in the detector using GEANT4 \cite{geant,atlas_geant}, and processed through the same reconstruction software as the data. The efficiency for charged hadrons is found to depend only weakly on the event multiplicity and on $p_{\mathrm{T}}$ for transverse momenta above 0.5~GeV. An efficiency of about 82\% is observed at mid-rapidity, $|\eta|<1$, decreasing to about 68\% at $|\eta|>2$. For low-\pT\ tracks, between 0.3~GeV and 0.5~GeV, the efficiency ranges from 74\% at $\eta= 0$ to about 50\% for $|\eta|>2$. The number of reconstructed charged particle tracks, not corrected for tracking efficiency, is denoted by $N_{\mathrm{ch}}^{\mathrm{rec}}$.

The analysis is performed in different intervals of $\Sigma E^{\mathrm{Pb}}_{\mathrm{T}}$, the sum of transverse energy measured in the FCal with $3.1 < \eta < 4.9$ in the direction of the Pb beam with no correction for the difference in response to electrons and hadrons.  The distribution of $\Sigma E^{\mathrm{Pb}}_{\mathrm{T}}$ for events passing all selection criteria is shown in Fig.~\ref{fig:EtNtrk}. These events are divided into six $\Sigma E^{\mathrm{Pb}}_{\mathrm{T}}$ intervals to study the variation of $v_2$ with overall event activity, as indicated in Fig.~\ref{fig:EtNtrk} and shown in Table~\ref{tab:centET}. Event ``activity'' is characterized by $\Sigma E^{\mathrm{Pb}}_{\mathrm{T}}$: the most active events are those with the largest $\Sigma E^{\mathrm{Pb}}_{\mathrm{T}}$. The distribution of $N_{\mathrm{ch}}^{\mathrm{rec}}$ for each activity interval is shown in the lower plot of Fig.~\ref{fig:EtNtrk}.

\section{Data analysis}
The cumulant method involves the calculation of $2k$-particle azimuthal correlations, $corr_n\{2k\}$,  and cumulants, $c_n\{2k\}$, where $k=1,2$ for the analysis presented in this paper. The two- and four-particle correlations are defined as ${corr_n\{2\} = \langle \mathrm{e}^{\mathrm{i}n(\phi_1 -\phi_2)} \rangle}$ and ${corr_n\{4\} = \langle \mathrm{e}^{\mathrm{i}n(\phi_1 +\phi_2-\phi_3-\phi_4)} \rangle}$, respectively, where the angle brackets denote the average in a single event over all pairs and all combinations of four particles. After averaging over events, the two-particle cumulant is obtained as ${c_n\{2\} = \langle corr_n\{2\} \rangle}$, and the four-particle cumulant ${c_n\{4\}=\langle corr_n\{4\} \rangle - 2 \cdot \langle corr_n\{2\} \rangle^2}$. Thus the effect of two-particle correlations is explicitly removed in the expression for $c_n\{4\}$. Further details are given in Refs.~~\cite{Borghini_first,Borghini_second,QC}.

Direct calculation of multi-particle correlations requires multiple passes over the particles in an event, and requires extensive computing time in high-multiplicity events.  To mitigate this, it has been proposed in Ref.~\cite{QC} to express multi-particle correlations in terms of the moments of the flow vector $Q_n$, defined as $Q_{n} = \sum_{i} \mathrm{e}^{\mathrm{i}n\phi_i}$, where the index $n$ denotes the flow harmonic  and the sum runs over all particles in an event. This analysis is restricted to the second harmonic coefficient, $n=2$. The method based on the flow-vector moments enables the calculation of multi-particle cumulants in a single pass over the full set of particles in each event.

The cumulant method involves two main steps \cite{Borghini_first,Borghini_second}. In the first step, the so-called  ``reference'' flow harmonic coefficients are calculated using multi-particle cumulants for particles selected inclusively from a broad range in $p_{\mathrm{T}}$ and $\eta$ as:
 \begin{eqnarray}
&&  \label{eq:vn2ref} v_2^{\mathrm{ref}}\{2\}=\sqrt{c_2\{2\}}, \\
&&  \label{eq:vn4ref} v_2^{\mathrm{ref}}\{4\}=\sqrt[4]{-c_2\{4\}},
  \end{eqnarray}
where $v_2^{\mathrm{ref}}\{2\}$ ($v_2^{\mathrm{ref}}\{4\}$) denotes the reference estimate of the second-order anisotropy parameter obtained using two-particle, $c_2\{2\}$ (four-particle, $c_2\{4\}$) cumulants.

The flow-vector method is easiest to apply when the detector acceptance is azimuthally uniform \cite{QC}. A correction for any azimuthal non-uniformity in the reconstruction of charged particle tracks is obtained from the data \cite{flowReview}, based on an $\eta$--$\phi$ map of all reconstructed tracks. For each small $(\delta\eta=0.1,\delta\phi=2\pi/64)$ bin (labelled $i$), a weight is calculated as $w_i(\eta,\phi)=\langle N(\delta\eta) \rangle/N_i(\delta\eta,\delta\phi)$, where $\langle N(\delta\eta) \rangle$ is the event-averaged number of tracks in the $\delta\eta$ slice to which this bin belongs, while $N_i(\delta\eta,\delta\phi)$ is the number of tracks in an event within this bin. Using this weight forces the azimuthal angle distribution of reference particles to be uniform in $\phi$, but it does not change the $\eta$ distribution of reconstructed tracks. A weighted $Q$-vector is evaluated as $Q_{n} = \sum_{i}w_i \mathrm{e}^{\mathrm{i}n\phi_i}$ \cite{QC}.
From Eqs.~(\ref{eq:vn2ref}) and (\ref{eq:vn4ref}) it is clear that the cumulant method can be used to estimate $v_2$ only when $c_2\{4\}$ is negative and $c_2\{2\}$ positive.

In the second step, the harmonic coefficients are determined as functions of  \pT\ and $\eta$, in bins in each variable (10 bins of equal width are used in $\eta$ and 22 bins of varied width in \pT). These differential flow harmonics are calculated for ``particles of interest'' which fall into these small bins. First, the differential cumulants, $d_2\{2\}$ and $d_2\{4\}$, are obtained by correlating every particle of interest with one and three reference particles respectively. The differential second harmonic, $v_2\{2k\}(\pT,\eta)$, where $k=1,2$, is then calculated with respect to the reference flow as derived in Refs.~\cite{Borghini_first,Borghini_second}:
 \begin{eqnarray}
&&  \label{eq:dvn2d} v_2\{2\}(p_{\mathrm{T}},\eta)=\frac{d_2\{2\}}{\sqrt{c_2\{2\}}},\\
&&  \label{eq:dvn4d} v_2\{4\}(p_{\mathrm{T}},\eta)=\frac{-d_2\{4\}}{\sqrt[3/4]{-c_2\{4\}}}.
  \end{eqnarray}

The differential $v_2$ harmonic is then integrated over wider phase-space bins, with each small bin weighted by the appropriate charged particle multiplicity. This is obtained from the reconstructed multiplicity by applying $\eta$- and \pT-dependent efficiency factors, determined from Monte Carlo (MC) simulation as discussed in the previous section.
%for weighting by the efficiency corrected charged particle multiplicity.
Due to the small number of events in the data sample, the final results are integrated over the full acceptance in $\eta$.
\begin{figure}[h!]
    \begin{center}
    \includegraphics[width=65mm]{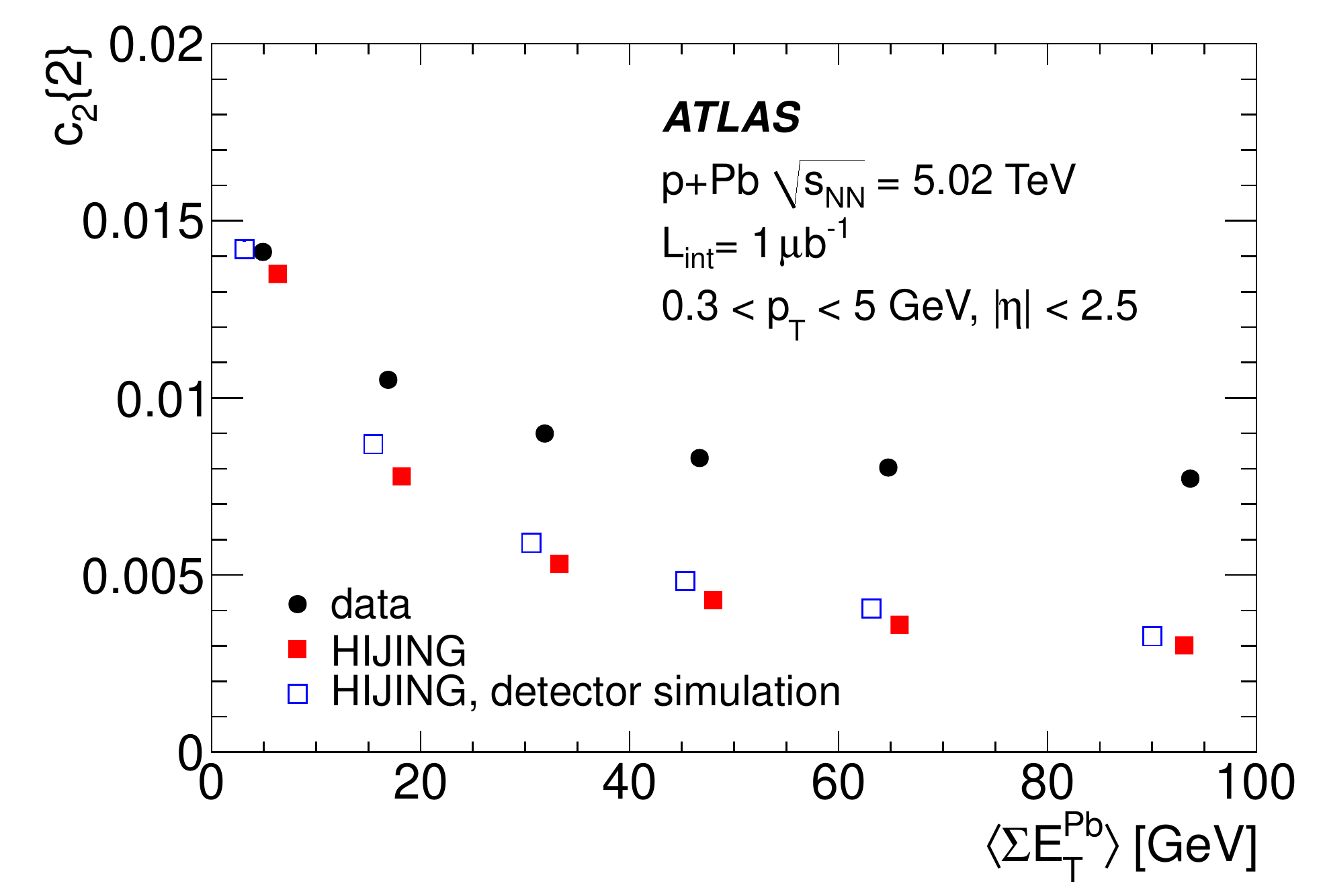}
    \includegraphics[width=65mm]{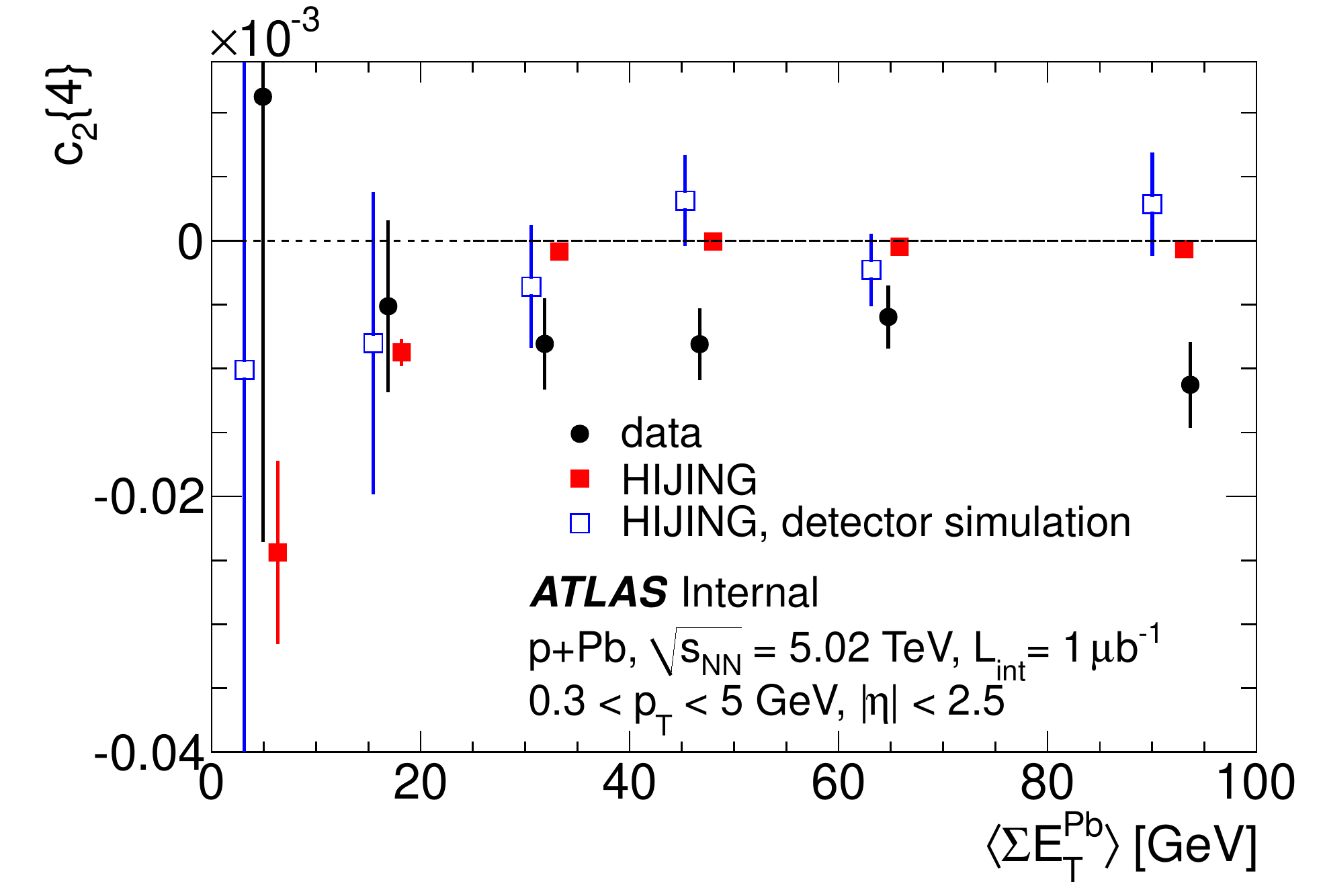}
  \caption{The two-particle (upper plot) and four-particle (lower plot) cumulants calculated using the reference flow particles as a function of $\Sigma E^{\mathrm{Pb}}_{\mathrm{T}}$ for data (circles), the fully simulated HIJING events (open squares) and the large generator-level HIJING sample (filled squares). For clarity, the points for the fully simulated (generated) HIJING events are slightly shifted to the left (right).}
  \label{fig:cnRef}
  \end{center}
\end{figure}

Fig.~\ref{fig:cnRef} shows the two- and four-particle cumulants, averaged over events in each event-activity class defined in Table~\ref{tab:centET}, as a function of $\Sigma E^{\mathrm{Pb}}_{\mathrm{T}}$.
It is observed that four-particle cumulants are negative  only in a certain range of event activity. This restricts subsequent analysis to events with $\Sigma E^{\mathrm{Pb}}_{\mathrm{T}}>25$~GeV, for which the four-particle cumulant in data is found to be less than zero by at least two standard deviations (statistical errors only). It was also checked that for these events $c_2\{4\}$ is unchanged within errors for any high-multiplicity selection. For example, defining $N_{20}$ as the value of $N_{\mathrm{ch}}^{\mathrm{rec}}$  such that 20\% of events have $N_{\mathrm{ch}}^{\mathrm{rec}} < N_{20}$ (i.e. $N_{20}$ is the 20th percentile), then selecting $N_{\mathrm{ch}}^{\mathrm{rec}} > N_{20}$ leaves $c_2\{4\}$  unchanged within errors. And for $\Sigma E^{\mathrm{Pb}}_{\mathrm{T}}>25$~GeV this holds for any percentile selection.

Fig.~\ref{fig:cnRef} also shows  the cumulants calculated for 50 million HIJING-generated events, using the true particle information only, as well as for one million fully simulated and reconstructed HIJING events, using the same methods as used for the data. The $\Sigma E^{\mathrm{Pb}}_{\mathrm{T}}$ obtained from the HIJING sample is rescaled to match that measured in the data. It should be noted that the HIJING Monte Carlo model does not contain any collective flow, and the only correlations are those due to resonance decays, jet production and momentum conservation.
The values of $c_2\{2\}$ for HIJING events are smaller than the values obtained from the data, and there is no significant difference between the HIJING results obtained at the generator (``truth'') level and at the reconstruction level.
\begin{figure*}[t]
  \begin{center}
    \includegraphics[width=160mm]{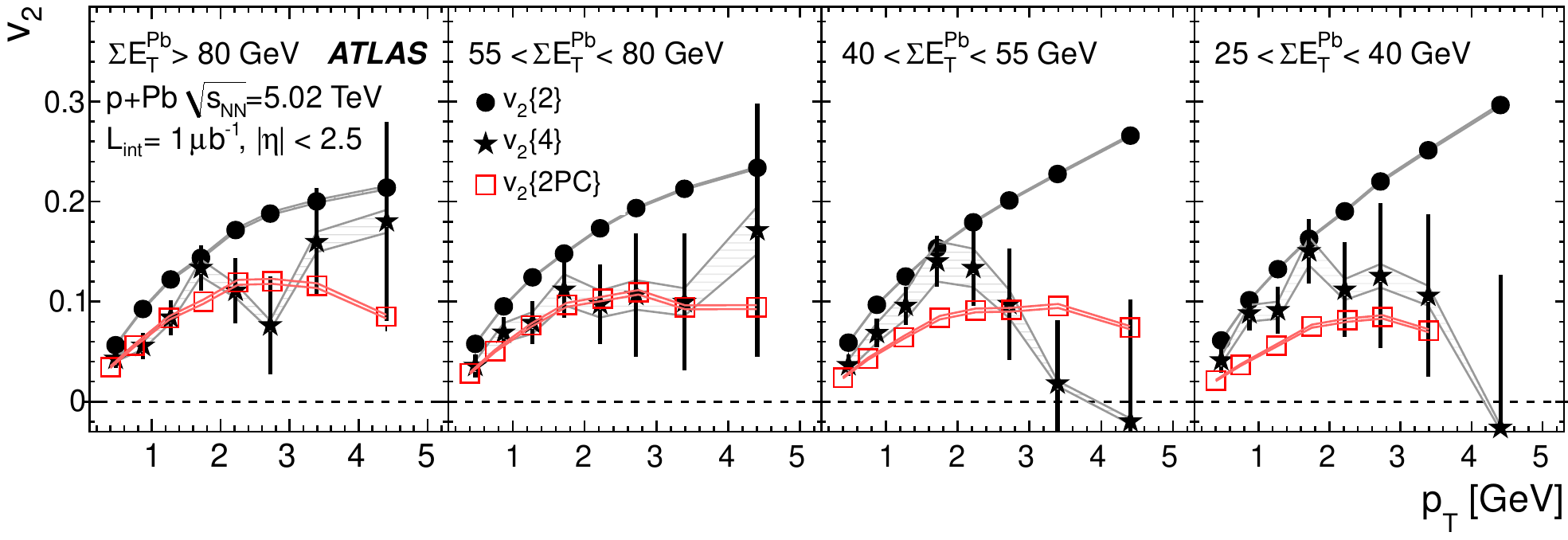}
    \caption{The second-order harmonic calculated with the two-particle (circles) and four-particle (stars) cumulants as a function of transverse momentum in four different activity intervals. Bars denote statistical errors; systematic uncertainties are shown as shaded bands. The $v_2$ derived from the Fourier decomposition of two-particle correlations \cite{ridge_atlas} is shown by squares.}
    \label{fig:v2pT}
  \end{center}
\end{figure*}
For $c_2\{4\}$, the HIJING events at $\Sigma E^{\mathrm{Pb}}_{\mathrm{T}}\sim 20$~GeV show a negative value comparable to the values seen in the data, indicating that correlations from jets or momentum conservation contribute significantly to $v_2\{4\}$ in events of low multiplicity. For $\Sigma E^{\mathrm{Pb}}_{\mathrm{T}}>25$~GeV the generator-level HIJING sample's values for $c_2\{4\}$ are also negative, but the magnitude is much smaller than in the data or in HIJING events with smaller $\Sigma E^{\mathrm{Pb}}_{\mathrm{T}}$. The size of the fully simulated HIJING event sample is too small to draw a definite conclusion about the sign or magnitude of $c_2\{4\}$.
 
The systematic uncertainties on $v_2\{2\}$ and $v_2\{4\}$ as a function of \pT\ and $\Sigma E^{\mathrm{Pb}}_{\mathrm{T}}$ have been evaluated by varying several aspects of the analysis procedure.
Azimuthal-angle sine terms in the Fourier expansion should be zero, but a non-zero contribution can arise due to detector biases. It was found that the magnitude of the sine terms relative to the cosine terms is negligible (below 1\%) for $v_2\{2\}$ measured as a function of \pT, as well as for the \pT-integrated $v_2\{2\}$ and $v_2\{4\}$. In the case of the measurement of the \pT-dependent $v_2\{4\}$, the systematic uncertainty attributed to the residual sine terms varies between 6\% and 14\% in the different $\Sigma E^{\mathrm{Pb}}_{\mathrm{T}}$ intervals. Uncertainties related to the tracking are obtained from the differences between the main results and those using tracking requirements modified to be either more or less restrictive. They are found to be negligible (below 0.2\%) for $v_2\{2\}$. For the \pT-dependent $v_2\{4\}$ they give a contribution of less than 6\% to the systematic uncertainty, and less than 1\% for the \pT-integrated $v_2\{4\}$. In addition to varying the track quality requirements, an uncertainty on the \pT\ dependence of the efficiency corrections is also taken into account, and found to be below 1\% for the $v_2\{2\}$ and $v_2\{4\}$ measurements. The correction of the azimuthal-angle uniformity is checked by comparing the results to those obtained with all weights, $w_i$, set equal to one. This change leads to small relative differences, below 1\%, for the $v_2\{2\}$ measured as a function of \pT, as well as for the \pT-integrated $v_2\{2\}$ and $v_2\{4\}$. Up to 4\% differences are observed in the \pT-dependent $v_2\{4\}$.  All individual contributions to the systematic uncertainty are added in quadrature and quoted as the total systematic uncertainty. The total systematic uncertainties are below 1\% for the $v_2\{2\}$ measurement. The $v_2\{4\}$ measurement precision is limited by large statistical errors, whereas the systematic uncertainties stay below 15\% for $v_2\{4\}(\pT)$ and below 2\% for the \pT-integrated $v_2\{4\}$.

\section{Results}
Fig.~\ref{fig:v2pT} shows the transverse momentum dependence of $v_2\{2\}$ and $v_2\{4\}$ in four different classes of the event activity, selected according to $\Sigma E^{\mathrm{Pb}}_{\mathrm{T}}$. A significant second-order harmonic is observed. $v_2\{4\}$ is systematically smaller than $v_2\{2\}$, consistent with the suppression of non-flow effects in $v_2\{4\}$. This difference is most pronounced at high \pT\ and in collisions with low $\Sigma E^{\mathrm{Pb}}_{\mathrm{T}}$ where jet-like correlations not diluted by the underlying event can contribute significantly. Thus, $v_2\{4\}$ appears to provide a more reliable estimate of the second-order anisotropy parameter of collective flow. As a function of transverse momentum the second-order harmonic, $v_2\{4\}$, increases with $p_{\mathrm{T}}$ up to $p_{\mathrm{T}}\approx$ 2~GeV. Large statistical errors preclude a definite conclusion about the \pT\ dependence of $v_2\{4\}$ at higher transverse momenta.

The shape and magnitude of the \pT-dependence of $v_2\{4\}$ is found to be similar to that observed in $p+$Pb collisions using two-particle correlations \cite{alice_ridge,ridge_atlas}. The second-order harmonic, $v_2$,  can be extracted  from two-particle azimuthal correlations using charged particle pairs with a large pseudorapidity gap to suppress the short-range correlations on the near-side ($\Delta\phi\sim 0$) \cite{ridge_atlas,atlas_2p}. However, the two-particle correlation measured this way may still be affected by the dijet correlations on the away-side ($\Delta\phi \sim \pi$), which can span a large range in $\Delta\eta$. In Ref.~\cite{ridge_atlas}, the away-side non-flow correlation is estimated using the yield measured in the lowest $\Sigma E^{\mathrm{Pb}}_{\mathrm{T}}$ collisions and is then subtracted from the higher $\Sigma E^{\mathrm{Pb}}_{\mathrm{T}}$ collisions. The result of that study, $v_2\{2PC\}$, is shown in Fig.~\ref{fig:v2pT} for the four activity intervals with largest $\Sigma E^{\mathrm{Pb}}_{\mathrm{T}}$,  and compared to $v_2\{4\}$. Good agreement is observed between $v_2\{4\}$ and $v_2\{2PC\}$ for collisions with $\Sigma E^{\mathrm{Pb}}_{\mathrm{T}}> 55$~GeV. For $\Sigma E^{\mathrm{Pb}}_{\mathrm{T}}< 55$~GeV, the disagreement could be due either to the subtraction procedure used to obtain $v_2\{2PC\}$ or to non-flow effects in $v_2\{4\}$, or to a combination.
\begin{figure}[h!]
  \begin{center}
    \includegraphics[width=75mm]{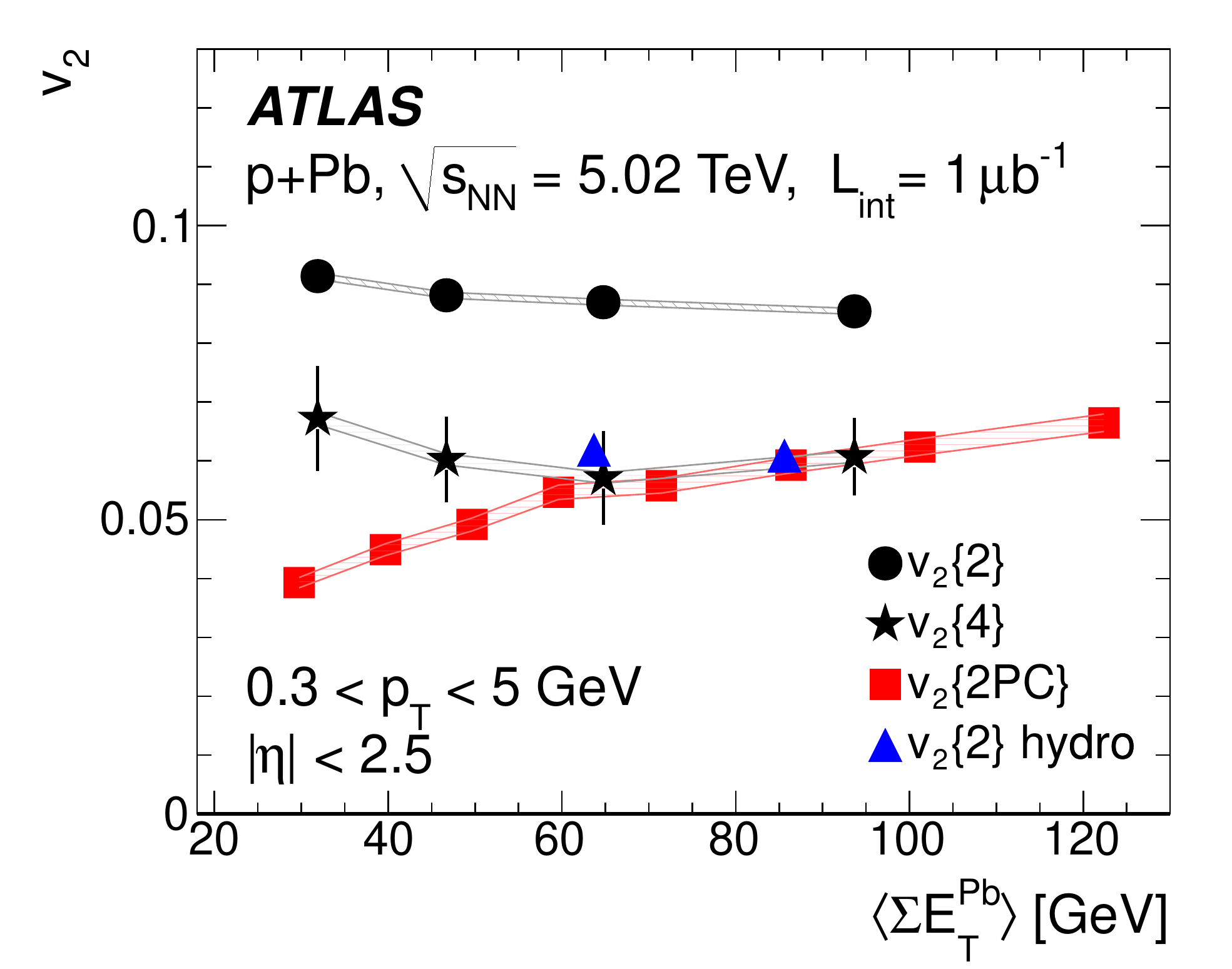}
    \caption{The second-order harmonic, $v_2$, integrated over $\pT $ and $\eta$, calculated with two- and four-particle cumulants (circles and stars, respectively), as a function  of $\Sigma E^{\mathrm{Pb}}_{\mathrm{T}}$. Systematic uncertainties are shown as shaded bands. Also shown is $v_2\{2PC\}$ (squares) and predictions from the hydrodynamic model \cite{bozek2} (triangles) for the same selection of charged particles as in the data. }
    \label{fig:v2Int}
  \end{center}
\end{figure}

The dependence on the collision activity of the second-order harmonic, integrated over  $0.3 < p_{\mathrm{T}} < 5$~GeV, is shown in Fig.~\ref{fig:v2Int}. The large magnitude of $v_2\{2\}$ compared to $v_2\{4\}$ suggests a substantial contamination from non-flow correlations. The value of $v_2\{4\}$ is approximately 0.06, with little dependence on the overall event activity for $\Sigma E^{\mathrm{Pb}}_{\mathrm{T}}>$25~GeV.  The extracted values of $v_2\{4\}$ are also compared to the $v_2\{2PC\}$ values obtained from two-particle correlations. Good agreement is observed at large $\Sigma E^{\mathrm{Pb}}_{\mathrm{T}}$, while at lower $\Sigma E^{\mathrm{Pb}}_{\mathrm{T}}$ the $v_2\{2PC\}$ is smaller than $v_2\{4\}$, which may be due to different sensitivity of the two methods to non-flow contributions that become more important in low $\Sigma E^{\mathrm{Pb}}_{\mathrm{T}}$ collisions.
Although $v_2\{4\}$ is constructed to suppress local two-particle correlations, it may still include true multi-particle correlations from jets, which should account  for a larger fraction of the correlated particle production in the events with the lowest $\Sigma E^{\mathrm{Pb}}_{\mathrm{T}}$.
If the HIJING results, shown in Fig.~\ref{fig:cnRef}, were used to correct the measured cumulants for this non-flow contribution, the extracted  $v_2\{4\}$ would be decreased by at most 10\% for $v_2\{4\}$ shown in Fig.~\ref{fig:v2Int}.  However, this correction is not applied to the final results.

It is notable that the trend of the \pT\ dependence of both $v_2\{4\}$ and $v_2\{2PC\}$ in $p+$Pb collisions resembles that observed for $v_2$ measured with the event-plane method in Pb+Pb collisions at $\sqrt{s_{\mathrm{NN}}}=2.76$~\TeV\ \cite{Atlas1,atlas_2p}, although with a magnitude between that observed in the most central and peripheral Pb+Pb collisions.  While the trend is found to be nearly independent of the Pb+Pb collision geometry, the magnitude in Pb+Pb events depends on the initial shape of the colliding system, and has been modelled for $\pT<2$~ GeV using viscous hydrodynamics \cite{luzum,gale,heinz}.

Harmonic flow coefficients in $p+$Pb collisions at  $\sqrt{s_{\mathrm{NN}}}=5.02$~\TeV\ have also been predicted using viscous hydrodynamics, with similar initial conditions as the Pb+Pb calculations \cite{bozek2}. The predicted magnitude of the second-order harmonic\footnote{We are grateful to the authors of Ref.~\cite{bozek2} for providing us with the model predictions for charged particles with $\eta$ and \pT\ ranges matching those used in the analysis. The model predictions are for two activity intervals corresponding to fractions of events of $0 - 3.4$ \% and $3.4 - 7.8$ \% which are then translated into the $\Sigma E^{\mathrm{Pb}}_{\mathrm{T}}$ intervals using Fig.~\ref{fig:EtNtrk}.} is compared to the measured $v_2\{4\}$ and $v_2\{2PC\}$ in Fig.~\ref{fig:v2Int}. It can be seen that the hydrodynamic predictions agree with our measurements over the $\Sigma E^{\mathrm{Pb}}_{\mathrm{T}}$ range where the model predictions are available.

\section{Conclusions}
ATLAS has measured the second  harmonic coefficient in $p+$Pb collisions at $\sqrt{s_{\mathrm{NN}}}=5.02$~\TeV\ using two- and four-particle cumulants. A significant magnitude of $v_2$ is observed using both two- and four-particle cumulants, although $v_2\{2\}$ is consistently larger than $v_2\{4\}$, indicating a sizeable contribution of non-flow correlations to $v_2\{2\}$. The transverse momentum dependence of $v_2\{4\}$ shows a behaviour similar to that measured in Pb+Pb collisions at $\sqrt{s_{\mathrm{NN}}}=2.76$~\TeV. The magnitude of $v_2\{4\}$ increases with \pT\ up to about 2--3~GeV. As a function of the collision activity, $v_2\{4\}$ remains constant, at the level of about 0.06, for the collisions with $\Sigma E^{\mathrm{Pb}}_{\mathrm{T}}>$25~GeV, which corresponds to about 40\% of the data. The measured $v_2\{4\}$ is found to be consistent with the second harmonic coefficient extracted by the Fourier decomposition of the long-range two-particle correlation function for collisions with $\Sigma E^{\mathrm{Pb}}_{\mathrm{T}}>55$~GeV. Good agreement is also found with the predictions of a hydrodynamic calculation for $p+$Pb collisions.

Extending previous results based only on two-particle correlations, the multi-particle cumulant results presented here provide additional evidence for the importance of final-state effects in the highest multiplicity $p+$Pb reactions. Final-state effects may lead to collective flow similar to that observed in the hot, dense system created in high-energy heavy-ion collisions.

% Acknowledgements for papers with collision data
% Version 5-Nov-2012

\section{Acknowledgements}

% Standard acknowledgements start here
%----------------------------------------------
We thank CERN for the very successful operation of the LHC, as well as the
support staff from our institutions without whom ATLAS could not be
operated efficiently.

We acknowledge the support of ANPCyT, Argentina; YerPhI, Armenia; ARC,
Australia; BMWF and FWF, Austria; ANAS, Azerbaijan; SSTC, Belarus; CNPq and FAPESP,
Brazil; NSERC, NRC and CFI, Canada; CERN; CONICYT, Chile; CAS, MOST and NSFC,
China; COLCIENCIAS, Colombia; MSMT CR, MPO CR and VSC CR, Czech Republic;
DNRF, DNSRC and Lundbeck Foundation, Denmark; EPLANET, ERC and NSRF, European Union;
IN2P3-CNRS, CEA-DSM/IRFU, France; GNSF, Georgia; BMBF, DFG, HGF, MPG and AvH
Foundation, Germany; GSRT and NSRF, Greece; ISF, MINERVA, GIF, DIP and Benoziyo Center,
Israel; INFN, Italy; MEXT and JSPS, Japan; CNRST, Morocco; FOM and NWO,
Netherlands; BRF and RCN, Norway; MNiSW, Poland; GRICES and FCT, Portugal; MERYS
(MECTS), Romania; MES of Russia and ROSATOM, Russian Federation; JINR; MSTD,
Serbia; MSSR, Slovakia; ARRS and MVZT, Slovenia; DST/NRF, South Africa;
MICINN, Spain; SRC and Wallenberg Foundation, Sweden; SER, SNSF and Cantons of
Bern and Geneva, Switzerland; NSC, Taiwan; TAEK, Turkey; STFC, the Royal
Society and Leverhulme Trust, United Kingdom; DOE and NSF, United States of
America.

The crucial computing support from all WLCG partners is acknowledged
gratefully, in particular from CERN and the ATLAS Tier-1 facilities at
TRIUMF (Canada), NDGF (Denmark, Norway, Sweden), CC-IN2P3 (France),
KIT/GridKA (Germany), INFN-CNAF (Italy), NL-T1 (Netherlands), PIC (Spain),
ASGC (Taiwan), RAL (UK) and BNL (USA) and in the Tier-2 facilities
worldwide.
%----------------------------------------------

\bibliographystyle{elsarticle-num}

\onecolumn 
\clearpage 
% ATLAS Collaboration author list
% Data extracted on 12-Feb-2013 for paper reference HION-2013-01
%\documentclass[11pt]{article}
%\usepackage{a4wide}\begin{document}
\begin{flushleft}
{\Large The ATLAS Collaboration}

\bigskip

G.~Aad$^{\rm 48}$,
T.~Abajyan$^{\rm 21}$,
B.~Abbott$^{\rm 111}$,
J.~Abdallah$^{\rm 12}$,
S.~Abdel~Khalek$^{\rm 115}$,
A.A.~Abdelalim$^{\rm 49}$,
O.~Abdinov$^{\rm 11}$,
R.~Aben$^{\rm 105}$,
B.~Abi$^{\rm 112}$,
M.~Abolins$^{\rm 88}$
% ...
\bigskip

% ...
$^{11}$ Institute of Physics, Azerbaijan Academy of Sciences, Baku, Azerbaijan\\
$^{12}$ Institut de F\'{i}sica d'Altes Energies and Departament de F\'{i}sica de la Universitat Aut\`{o}noma de Barcelona and ICREA, Barcelona, Spain\\
% ...
$^{21}$ Physikalisches Institut, University of Bonn, Bonn, Germany\\
% ...
$^{48}$ Fakult\"{a}t f\"{u}r Mathematik und Physik, Albert-Ludwigs-Universit\"{a}t, Freiburg, Germany\\
$^{49}$ Section de Physique, Universit\'{e} de Gen\`{e}ve, Geneva, Switzerland\\
% ...
$^{88}$ Department of Physics and Astronomy, Michigan State University, East Lansing MI, United States of America\\
% ...
$^{105}$ Nikhef National Institute for Subatomic Physics and University of Amsterdam, Amsterdam, Netherlands\\
% ...
$^{111}$ Homer L. Dodge Department of Physics and Astronomy, University of Oklahoma, Norman OK, United States of America\\
$^{112}$ Department of Physics, Oklahoma State University, Stillwater OK, United States of America\\
% ...
$^{115}$ LAL, Universit\'{e} Paris-Sud and CNRS/IN2P3, Orsay, France\\
% ...
\end{flushleft}

%\end{document}
% Created with TexConverter
 
\clearpage

\end{document}